\documentclass[10pt,aps,prd,nofootinbib,superscriptaddress,twocolumn,preprintnumbers,balancelastpage]{revtex4}
\usepackage{amssymb,amsmath,latexsym,graphics, graphicx,epsfig,multirow,comment,appendix,feyn,slashed,xcolor,afterpage,multirow,makecell} 
\usepackage[colorlinks=true
,urlcolor=blue
,anchorcolor=blue
,citecolor=blue
,filecolor=blue
,linkcolor=blue
,menucolor=blue
,linktocpage=true
,pdfproducer=medialab
,pdfa=true
]{hyperref}
\usepackage[utf8]{inputenc}

\newcommand {\be} {\begin {equation}}
\newcommand {\ee} {\end {equation}}

\newcommand {\bes} {\begin {equation*}}
\newcommand {\ees} {\end {equation*}}

\def\CR{{\cal R}}

\def\CO{{\cal O}}

\newcommand{\beq}{\begin{equation}}
\newcommand{\eeq}{\end{equation}}

\newcommand{\Hz}{\,\textrm{Hz}}
\newcommand{\kHz}{\,\textrm{kHz}}
\newcommand{\MHz}{\,\textrm{MHz}}

\newcommand{\alpNP}{\alpha_{\rm NP}}
\newcommand{\Uhat}{ \vec{m\mu}}
\newcommand{\ambit}{\textsc{amb}i\textsc{t}}

\newcommand{\mnvec}{\vec{m\nu}}
\newcommand{\mrvec}{ \vec{m \delta \langle r^2 \rangle}}
\newcommand{\delr}[1]{\delta\langle r^2\rangle_{#1}}

\newcommand{\NL}{{\rm NL}}

\begin{document}

%%%%%%%%%%%%%%%%%%%%%%%%%%%%%%%
\title{Generalized King linearity and new physics searches with isotope shifts}
%%%%%%%%%%%%%%%%%%%%%%%%%%%%%%%

\preprint{LAPTH-020/20}

%%%%%%%%%%%%%%%%%%%%%%%%%%%%%%%
\author{Julian C. Berengut}
\email{julian.berengut@unsw.edu.au}
\affiliation{School of Physics, University of New South Wales, Sydney, New South Wales 2052, Australia}

\author{C\'edric Delaunay}
\email{cedric.delaunay@lapth.cnrs.fr}
\affiliation{Laboratoire d'Annecy-le-Vieux de Physique Th\'eorique LAPTh, CNRS -- Universit\'e Savoie Mont Blanc, BP 110, F-74941 Annecy-le-Vieux, France}

\author{Amy Geddes}
\email{a.geddes@unsw.edu.au}
\affiliation{School of Physics, University of New South Wales, Sydney, New South Wales 2052, Australia}

\author{Yotam Soreq}
\email{soreqy@physics.technion.ac.il}
\affiliation{Physics Department, Technion---Israel Institute of Technology, Haifa 3200003, Israel}
%%%%%%%%%%%%%%%%%%%%%%%%%%%%%%%

%\date{\today}

\begin{abstract}
Atomic spectral lines for different isotopes are shifted, revealing a change in the properties of the nucleus. 
For spinless nuclei such isotope shifts for two distinct transitions are expected to be linearly related, at least at leading order in a change of the nuclear mass and charge distribution. 
Looking for a breaking of linearity in so-called King plots was proposed as a novel method to search for physics beyond the standard model. 
In the light of the recent experimental progress in isotope shift spectroscopy, the sensitivity of these searches will become limited by the determination of the isotope masses and/or by nuclear effects which may induce nonlinearities at an observable level.
In this work, we propose two possible generalizations of the traditional King plot that overcome these limitations by including additional isotope shift measurements, thus significantly extending the new physics reach of King plots in a purely spectroscopy-driven approach. 
\end{abstract}

\maketitle

%%%%%%%%%%%%%%%%%%%%%%%%%%%%%%%
\section{Introduction} 
\label{sec:intro}
%%%%%%%%%%%%%%%%%%%%%%%%%%%%%%%

Fundamental interactions of the known elementary particles are well described by the standard model~(SM) of particle physics. 
This theory has been tested up to TeV-scale energies by the ATLAS and CMS experiments conducted at the Large Hadron Collider~(LHC). 

Despite its great experimental success, the SM cannot be a complete theory of Nature. First of all, it does not account for observations related to the matter-antimatter asymmetry of the universe, neutrino oscillations and dark matter. 
Moreover, the SM seems incomplete also from a theoretical standpoint. 
Most importantly, it suffers from hierarchy problems related to the instability of the electroweak scale and the absence of CP violation in strong interactions. 
The SM also brings to light a deep flavor puzzle indicated by a very broad spectrum of elementary particle masses. 
All these considerations call for a completion of the SM, in the form of new interactions and particles. 

Unfortunately, none of the above shortcomings of the SM unarguably points towards a preferred theoretical construct, nor even a specific energy scale for such a new physics~(NP). For many decades, strongly motivated by the electroweak-scale instability problem, physics beyond the SM~(BSM) was believed to emerge only at or slightly above the TeV scale. Today, the lack of evidence from the LHC program for NP at high-energies challenges this paradigm. 
Interestingly, in the meantime, new BSM theories have been found where the aforementioned problems of the SM are solved by much lighter new particles, with mass well below the GeV scale and small couplings to the matter fields~\cite{Wilczek:1982rv,Gelmini:1982zz,Kim:1986ax,Feng:1997tn,Graham:2015cka, Gupta:2015uea, Flacke:2016szy, Frugiuele:2018coc,Banerjee:2020kww}. 

This wide window of theoretical possibilities greatly motivates broad searches for NP at very different energy scales. 
While the accelerator-based program searches for NP at the high-energy frontier, precision measurements at low energies form a highly valuable complementary approach. See Refs.~\cite{Ginges:2003qt,Karshenboim:2005iy,Safronova:2017xyt} for reviews. Low-energy measurements indirectly probe NP whose existence may be revealed either by comparing the SM predictions to the experimental results or by searching for signatures forbidden by (approximate)~symmetries of the SM.
Well known examples are the searches for electric dipole moments of elementary particles~\cite{Andreev:2018ayy,Baron:2013eja} or precision measurements of their magnetic dipole moments~\cite{Hanneke:2008tm} and that of atomic parity-violating transitions~\cite{Wood:1997zq,Guena:2004sq}.\\

Atomic spectral lines, and their isotope shifts~(IS), are also extremely well measured. 
Yet, their calculation is not possible with comparable accuracy, except in very simple atoms like hydrogen, deuterium and helium or in purely leptonic bound states like positronium or muonium (for related work see \textit{e.g.}~\cite{Frugiuele:2019drl,Pachucki:2017xcg,Karshenboim:2000kv}). 
For heavier elements, a promising approach~\cite{Delaunay:2016brc} consists in combining precision measurements of IS in so-called King plots~\cite{King:63} and testing for King linearity. 
In this approach, current IS spectroscopy may already probe new spin-independent interactions between electrons and neutrons at an unprecedented level~\cite{Berengut:2017zuo,Frugiuele:2016rii}. 
Moreover, the sensitivity in probing NP with IS is expected to greatly improve in the near future as indicated by recent experimental developments. 
For instance, IS spectroscopy reached an accuracy of better than 10\,Hz for electric quadrupole transitions in Ca$^+$ ions~\cite{Knollmann:2019gcc}. 
Also, a novel strategy of IS measurement was proposed~\cite{Manovitz:2019czu} which is based on preparing pairs of different isotopes in an entangled state that oscillates at the IS frequency. 
By significantly improving on noise reduction, this technique allowed IS measurements at the $10\,$mHz level. While this very high reach in accuracy was first demonstrated with Sr isotopes, it can in principle be implemented in other elements relatively easily.  

As the accuracy of IS measurements keeps on improving, the potential to probe NP with King linearity will be curbed by two possible limitations. 
First, the uncertainty associated with nuclear masses will eventually exceed the IS measurement uncertainty. 
The King plot analysis is constructed in terms of so-called modified IS, which are IS frequencies divided by the difference of (inverse) isotope masses. 
Therefore, nuclear masses must be known also with a very high precision. 
Typically, the relative mass uncertainty will become a limiting factor whenever it is comparable or larger than that of the IS measurement multiplied by the relative mass difference.  
For instance, assuming state-of-the-art mass determination~\cite{Audi_2017,Wang_2017}, mass uncertainties of ytterbium~(Yb) isotopes are expected to limit King linearity tests  once the precision in IS measurements goes below the Hz level.

The second limitation comes from nonlinearities~(NLs) of the King plot due to higher order nuclear corrections in the SM. 
Linearity is the result of an approximate factorization of the electronic and nuclear effects in the two dominant contributions to the IS.
This factorization is expected to break once sub-leading nuclear effects are included, which thus induces King linearity violation, namely NLs in King plots. 
Such NLs are expected to depend on the element and are challenging to calculate theoretically. 
Nevertheless, their size has been estimated in a few cases. For example, NLs were found to be rather small (below the Hz level) in Ca$^+$ but those could reach up to $\CO(10)\,\kHz$ in King plots based on Yb$^+$ clock transitions~\cite{Flambaum:2017onb} or fine-structure transitions in argon ions~\cite{Yerokhin:2020oes}, for example. 
With a state-of-the-art accuracy in IS measurements at or even below the Hz level for some of these transitions, subleading nuclear effects will soon become an important limitation in the search for NP. 

Very recently, dedicated NP searches with IS spectroscopy have been reported in Yb$^+$~\cite{Counts:2020aws} and Ca$^+$~\cite{Solaro:2020dxz}, which constitute a significant improvement upon the prior most accurate test of King linearity performed with dipole transitions in Ca$^+$ measured at a precision of $\CO(100)\,\kHz$~\cite{PhysRevLett.115.053003}.
The authors of Ref.~\cite{Counts:2020aws} constructed a King plot from two quadrupole transitions in Yb$^+$ ($^2S_{1/2}\!\to \! ^2D_{3/2}$ and $^2S_{1/2}\!\to \! ^2D_{5/2}$) for which IS were measured with an accuracy of $\CO(300)\,\Hz$.
The resulting King plot shows an evidence of NL at the three standard deviation level. 
Such NL may result from NP or higher order SM contributions.
With only ytterbium data, the two interpretations cannot be distinguished.
In addition, the current accuracy of these measurement also indicates that the uncertainty due to the isotope mass determination already plays a role in the sensitivity to NP with the King analysis.   
In the meantime, the authors of Ref.~\cite{Solaro:2020dxz} measured the $3d\ ^2D_{3/2}\!\to\!^2D_{5/2}$ fine-structure transition in Ca$^+$ with $\CO(20)\,\Hz$ accuracy and constructed a King plot by combining it with measurements of the $^2S_{1/2}\!\to\!^2D_{5/2}$ transition. The resulting King plot is consistent with linearity and has a comparable sensitivity to NP as the Yb$^+$ analysis of Ref.~\cite{Counts:2020aws}. Therefore, as illustrated in Fig~\ref{fig:Yb2D3DGen} below, the NP explanation of the Yb$^+$ data appears in tension with the Ca$^+$ data, thus favoring a SM origin for the non linear behaviour reported in Ref.~\cite{Counts:2020aws}.

In this paper we show how to overcome the limitations due to too large isotope mass uncertainties and/or higher order SM contributions without resorting to improved nuclear mass determination, nor theory calculations of nuclear NLs. 
Basically, we propose to use one (or more) additional IS transition(s) in order to eliminate the nuclear parameter(s) whose knowledge is not sufficiently accurate to allow for NP searches driven only by very precise IS spectroscopy.
Reference~\cite{Mikami:2017ynz} pointed out that it is possible to keep improving the NP sensitivity even in the presence of NLs by constructing a generalized King plot where the subleading nuclear physics parameters are absorbed by including IS measurements in three or more transitions.  
Here we further develop the generalized King plot analysis of Ref.~\cite{Mikami:2017ynz} and show the potential improvement sensitivity to NP beyond the leading order contributions. 

The rest of the paper is organized as follows.
Section~\ref{sec:generalking} contains a review of the King analysis.
In Section~\ref{sec:masslessking}, we present a modified King analysis without the input of isotope masses. 
A parameterization of higher order SM effects is presented Section~\ref{sec:generalkingNuclear}.
We show a generalization of the King analysis which include higher order effects in Section~\ref{sec:generalkingNP}.
We demonstrate the above two methods in Section~\ref{sec:Yb} by examination of the ytterbium case.
Finally, we conclude in Section~\ref{sec:summary}.

%%%%%%%%%%%%%%%%%%%%%%%%%%%%%%%
\section{Review of King linearity and its breaking} 
\label{sec:generalking}
%%%%%%%%%%%%%%%%%%%%%%%%%%%%%%%

Consider a set of atomic transitions $i$ and their IS frequencies $\nu_i^{a}\equiv\nu_i^{A_0}-\nu_i^A$ between two isotopes of mass number $A_0$ and $A$, with $a\equiv A_0 A$.\footnote{Here we consider $A_0$ as the reference isotope, however, it also possible to preform the King analysis with neighboring isotope pairs.} 
The latter is usually understood as being comprised of two SM terms~\cite{King:63} and possibly one NP term~\cite{Delaunay:2016brc}, 
\begin{align}
	\label{eq:LOshift}
	\nu_i^{a} 
= 	K_i \mu_{a} + F_i \delr{a}+ \alpNP X_i \gamma_{a} \,.
\end{align}
The first term is the mass shift~(MS), with $\mu_{a}\equiv 1/m_{A_0}-1/m_{A}$ being the difference of inverse nuclear masses, and the second term is the volume or field shift~(FS). 
The third term in Eq.~\eqref{eq:LOshift} is the result of NP, where $\alpNP$ denotes the NP coupling strength, while $X_i$ and $\gamma_a$ are the NP electronic constant and nuclear parameter, respectively.   
FS originates from differences in the nuclear charge distributions due to the different number of neutrons and, to a good approximation, is proportional to the difference of rms charge radii $\delr{a}\equiv \langle r^2\rangle_{A_0}-\langle r^2\rangle_{A}$. 
$K_i$, $F_i$ and $X_i$ are constants that only depend on the electronic configurations associated with the transition $i$. 
This factorization of electronic and nuclear parameters can be understood from perturbation theory by noting that the nuclear mass is much larger than the electron mass $m_e/m_A\ll 1$, and that only a small portion of the electron density penetrating the nucleus is sensitive to the finite nuclear size. 

Whenever Eq.~\eqref{eq:LOshift} is a valid description of the IS, it predicts that the so-called modified IS~(mIS), $m\nu_i^{a} \equiv \mu^{-1}_{a}\nu_i^{a}$, for two distinct transitions $i=1,2$ are linearly related~\cite{King:63} in the absence of NP, $\alpNP=0$. 
Indeed, the difference $\delr{a}$ can be traded for a measurement in one transition as $\mu_{a}^{-1}\delr{a}=(m\nu_1^{a}-K_1)/F_1$, yielding for the second transition $m\nu_2^{a} = F_{21} m\nu_1^{a} + K_{21}\,$, where $F_{21}\equiv F_2/F_1$ and $K_{21}\equiv K_2-F_{21}K_1$.  
This relation is known as King linearity. 
Note that knowledge of the nuclear charge distributions and the electronic constants $K_i$ and $F_i$ is not needed at all in order to make this prediction. 
It does, however, require determination of the nuclear masses with good accuracy. 

New atomic forces typically break King linearity~\cite{Delaunay:2016brc,Berengut:2017zuo}. 
In the presence of NP, $ \alpNP \ne 0$, and repeating the steps above gives
\begin{align}
	\label{eq:KL}
	m\nu_2^{a} 
= 	F_{21}m\nu_1^{a} + K_{21} +\alpNP X_{21} h_{a}\,,
\end{align}
where $X_{21}\equiv X_2-F_{21}X_1$ and $h_{a}\equiv \mu_{a}^{-1}\gamma_{a}$. 
The NP term thus induces NL, unless 
(i)~the ratio $X_i/F_i$ is constant between the two transitions or 
(ii)~$h_{a}$ is either independent of $a$ or $\propto m\nu_{1,2}^{a}$. 
The former arises when the range of the new force becomes shorter than the nuclear size. 
In this case, the atomic potentials for both FS and NP are described by a contact interaction and probe the same region of the electron cloud inside the nucleus, leading to $X_i\propto F_i$. 
Nevertheless, high sensitivity to the NP coupling $\alpNP$ can be achieved away from these limits~\cite{Berengut:2017zuo}. 
The NP coupling can be determined by solving Eq.~\eqref{eq:KL} for three different isotope pairs  $a_{1,2,3}$, yielding
\begin{align}
	\label{eq:alpNPIS}
	\alpNP
=	\frac{\det(\mnvec_1,\mnvec_2,\Uhat)}{\det(X_1 \mnvec_2-X_2\mnvec_1,\vec{h},\Uhat)} \, ,
\end{align}
where arrows denote vectors in the isotope space and $\Uhat=(1,1,1) $. 
Note that the electronic constants $X_{1,2}$ and the nuclear parameter $\vec{h}$ of NP are the only theoretical inputs needed to extract $\alpha_{\rm NP}$ given a set of $\mnvec_{1,2}$ data. 
Equation~\eqref{eq:alpNPIS} has a clear geometrical interpretation. 
The numerator on the right-hand side is the amount of NL in the data, which is related to the area of the triangle formed by three points on the King plot, while the denominator corresponds to the amount of NL predicted by theory (for a coupling of unity, $\alpNP=1$). 
Using Eq.~\eqref{eq:LOshift} the latter writes $(F_1X_2-F_2X_1)\times \det(\mrvec,\vec{h},\Uhat)$. 
The first factor shows the suppression due to the alignment of electronic FS and NP constants in the high mass limit. 
The second factor is the alignment of nuclear parameters. 
Note that for $\gamma_a=\Delta A=A_0-A$, which is typically the case for new spin-independent interactions, the NP and MS nuclear parameters are approximately aligned since $\vec{h} \simeq -A_0^2 (\Uhat - \vec{\Delta A}/A_0)$ resulting for instance in a sensitivity suppression of $\CO(20)$ for Yb.   
%

%%%%%%%%%%%%%%%%%%%%%%%%%%%%%%%
\section{King analysis without isotopes mass} 
\label{sec:masslessking}
%%%%%%%%%%%%%%%%%%%%%%%%%%%%%%%

The linear relation of Eq.~\eqref{eq:KL} (in the absence of NP) holds for $m\nu_i^a$, but not for $\nu_i^a$.
Therefore, testing King linearity requires good knowledge of the isotope masses. 
The total mIS uncertainty as a function of uncertainties in the IS and mass measurements is,
\begin{align}
	\label{eq:umIS}
	u[m\nu] 
= 	\sqrt{u[\nu]^2+\frac{m_{A}^2u[m_{A_0}]^2+m_{A_0}^2u[m_{A}]^2}{(m_{A_0}-m_{A})^2} }\,,
\end{align}
where  $u[x]\equiv \sigma[x]/x$, with $\sigma[x]$ denoting the standard deviation of observable $x$. 
Hence, isotope masses must be known better than IS by a factor $\CO(m_A/(m_{A_0} - m_{A}))$. 
Otherwise, the assessment of King linearity will be limited by the relatively poor knowledge of nuclear masses, resulting in a weaker sensitivity to NP forces. 

Here we present a simple modification of the King linearity definition that does not rely on precise isotope mass determination. 
The idea is to use IS measurements in a third transition (with the same isotope pairs) in order to remove the MS nuclear parameter that is not sufficiently well known.  
Consider the case of three transitions. 
The MS and FS nuclear parameters, respectively  $\mu_a$ and $\delr{a}$, are extracted using Eq.~\eqref{eq:LOshift} applied to two of the transitions, and then plugged in the theory prediction for the third one. 
This yields a linear relation among the IS of the three transitions, up to a NP contribution,
\begin{align}
	\label{eq:KingNonMass}
	\nu_3^{a} 
= 	f_3^\alpha \nu_\alpha^{a} 
	+\alpNP \left( X_3-f_3^\alpha X_\alpha \right) \gamma_{a}\,,
\end{align}
where sum over $\alpha=1,2$ is implicit and 
\begin{align}
	\left( f_3^1 , \,  f_3^2 \right) 
=	\left( K_3, \, F_3 \right) \cdot
	\begin{pmatrix}
		K_1 & F_1 \\
		K_2 & F_2
	\end{pmatrix}^{-1} \, .
\end{align}
Similarly to the usual King linearity, this prediction can be tested without any additional theory inputs by checking the consistency of $\det(\vec\nu_1,\vec\nu_2,\vec\nu_3)$ with zero. We refer to this method as the \textit{no-mass King~(NMK)} analysis.

In the presence of NP, we can extract $\alpNP$ by measuring IS of three transitions and three isotope pairs upon using Eq.~\eqref{eq:KingNonMass}, which yields
\begin{align}
	\label{eq:alphaNPmassless}
	\alpha_{\rm NP}
=	\frac{\det(\vec\nu_1,\vec\nu_2,\vec\nu_3)}{\frac{1}{2}\epsilon_{ijk}
	\det(\vec\nu_i,\vec\nu_j,X_k\vec\gamma)}
\end{align}
where the sum over the indices is implicit. 
As for the original King linearity case, the numerator of Eq.~\eqref{eq:alphaNPmassless} is a measure of NL present in the data, while the denominator corresponds to the amount of NL predicted by theory for $\alpNP=1$.  
Equation~\eqref{eq:LOshift} allows to express it in terms of the MS and FS parameters as $\epsilon_{ijk}K_iF_jX_k\times \det(\vec\mu,\vec{\delr{a}},\vec\gamma)$, which makes manifest the limits where sensitivity to NP is suppressed due to the alignment of either electronic constants (as in the heavy NP mass limit) or nuclear parameters. 
The NP sensitivity of the NMK analysis of Eq.~\eqref{eq:alphaNPmassless} and its numerical comparison with the original King plot analysis is discussed in Section~\ref{sec:Yb} for the case of Yb. 

%%%%%%%%%%%%%%%%%%%%%%%%%%%%%%%
\section{Nonlinearities in King Plots from nuclear effects} 
\label{sec:generalkingNuclear}
%%%%%%%%%%%%%%%%%%%%%%%%%%%%%%%

Equation~\eqref{eq:LOshift} is not a complete description of the IS, see \textit{e.g.}~\cite{king2013isotope}.
It is only approximate, valid at leading order in the variation of the nuclear mass and charge distribution. 
This is straightforward to understand from standard (time-independent) perturbation theory in quantum mechanics. 
At first order in a small perturbation (such as a change in nuclear parameters) only the eigenvalues of the Hamiltonian are corrected while the associated eigenstates are unchanged. 
Hence, at first order in the variation of nuclear mass and charge distribution, the electronic configurations remain the same between isotopes, yielding the factorized form in Eq.~\eqref{eq:LOshift}. 
However, at second order in the perturbation, the electron cloud will respond to a change of the nuclear properties and make both MS and FS electronic constants isotope dependent, $K_i\to K_i^a$ and $F_i\to F_i^a$, which induces King linearity breaking.

Experimentally, there are very few observations of King plot NLs. 
Large NLs of $\CO(10)\,\MHz$ were observed in samarium isotopes~\cite{Griffith_1979}, which however was later understood to result from a strong enhancement of second-order FS effect due to mixing with a nearly-degenerate level~\cite{Griffith_1981,Palmer_1982}. 
A nonlinear behaviour at the $100\,$kHz level was recently reported in strontium~\cite{miyake2019isotope}. 
However the King plot of Ref.~\cite{miyake2019isotope} was constructed using $^{87}$Sr which may indicate the presence of sizable effects from the nuclear spin of this odd isotope. Finally, we note that the very recent $3\sigma$ evidence for NL in Yb$^+$ quadrupole transitions~\cite{Counts:2020aws} is consistent with an interpretation in terms of second-order FS or may be an indication of NP.

Calculating the NLs induced by nuclear effects in a given element is theoretically challenging. 
The main difficulty arises from the fact that most of the higher order corrections are aligned with the leading order contributions in Eq.~\eqref{eq:LOshift} and therefore they do not yield NLs.   
A well understood example is higher moments in the nuclear charge distribution~\cite{SELTZER:1969zz}. 
The theoretical evaluation of NLs in King plots was reconsidered in Ref.~\cite{Flambaum:2017onb} which presented a first-time calculation of King plot NLs for narrow transitions in several heavy elements where the dominant contribution was found to arise from second order FS.  

With improvements in the experimental accuracy, additional nuclear parameters, such as higher moments of the charge distribution, may become relevant. 
As pointed out in Ref.~\cite{Mikami:2017ynz}\, these effects are formally parameterized, without loss of generality, by adding to Eq.~\eqref{eq:LOshift} an infinite series of terms involving new nuclear parameters different from $\mu_{a}$ and $\delr{a}$, 
\begin{align}
	\label{eq:ISgen}
	\nu_i^{a} 
=  	K_i\mu_{a}+ F_i \delr{a} + \sum_{l=2}^\infty F_{il} \lambda_{l,a}\,,
\end{align}
where $\lambda_{l,a}$ are independent nuclear parameters and $F_{il}$ are electronic constants. 
Once experimental precision in IS measurement reaches the level of the (dominant) sub-leading effects, SM NLs arise in King plots and it becomes challenging to probe NP. 

A possible strategy would be to calculate those NLs from atomic and/or nuclear theory, subtract them from the measurements, and construct a King plot with the residual mIS. 
However, as already emphasized, such calculations are rather challenging. 
Moreover, to understand the level at which the SM nonlinearity will be observed, the  calculation of an isotope invariant measure, like Eq.~\eqref{eq:2DNL} below, is needed.

Another strategy, where two transitions and four isotope pairs are available is presented in Ref.~\cite{Counts:2020aws}.
Instead of constructing a King plot of the two mIS, they considered the correlation between one transition and $\mu_a$, normalized by the second transition, which is predicted to be linear by the SM at leading order.
The leading SM source of NL and NP can be distinguished if prior knowledge about the former is included,  for instance if originating from second order FS, so that a particular pattern of NLs can be predicted. 
Then, provided  measurements (and theoretical inputs) are accurate enough, one can distinguish between the two effects.

%%%%%%%%%%%%%%%%%%%%%%%%%%%%%%%
\section{Generalized King linearity and New Physics} 
\label{sec:generalkingNP}
%%%%%%%%%%%%%%%%%%%%%%%%%%%%%%%

Here we further extend the generalized King plot analysis of Ref.~\cite{Mikami:2017ynz}, such that  physics beyond the SM could still be probed in the presence of NLs in King plots using minimal theoretical inputs. 
The idea is to use IS measurements of additional transitions in order to fix the nuclear and electronic parameters that break King linearity at leading order.

IS are small perturbations to the atomic spectra, with effects scaling like ratios of nuclear scales to atomic scales. 
Hence, the number of $\lambda_{l,a}$ that are relevant given a fixed experimental precision is expected to be finite and (hopefully) small. 
The exact number of such parameters and their nuclear structure, \textit{i.e.} their isotope dependence, could for instance be inferred from atomic theory calculations, see \textit{e.g.}~\cite{Flambaum:2017onb}. 

In the current analysis, we assume $l=2\dots m-1$ where $m$ is (at most) equal to the number clock transitions available for a given element, both neutral and charged counterparts.
Then, all mIS for that element depend on $m-1$ independent nuclear parameters, 
\begin{align}
	\label{eq:mnutheory}
	m\nu_i^{a}
= 	K_i+\sum_{l=1}^{m-1} F_{il}m\lambda_{l,a} + \alpha_{\rm NP} X_i h_{a}\,,
\end{align}
where for convenience we have defined $F_{i1}\equiv F_i$, $\lambda_{1,a}=\delr{a}$, and $m\lambda_{l,a}\equiv  \mu^{-1}_a\lambda_{l,a}$\,. 
Hence we need at least $m-1$ transitions to fix the unknown nuclear parameters
\begin{align}
	m\lambda_{l,a}
=  	(F^{-1})_{li} (m\nu_i^{a}-K_i-\alpha_{\rm NP} X_i h_{a})\,,
\end{align}
where $F^{-1}$ denotes the inverse of the matrix $F$ with $i,l=1\dots m-1$,  and there is an implied sum over $i$.
Consequently, the mIS for any other independent transition $k\neq 1\dots m-1$ is predicted to be linearly related to the previous $m-1$ ones, up to a linearity-breaking term from NP
\begin{align}
	\label{eq:linearrelation}
	m\nu_{k}^{a} 
\!=\! 	[K_k- f_{ki}K_i]\!+\! f_{ki}m\nu_i^{a}\!+\!\alpNP[X_{k}\!-\!f_{ki}X_i]h_{a},
\end{align}
where $k\ne i$ and $f_{ki}\! \equiv\! F_{kl}(F^{-1})_{li}$, where $l\ne k$ in the sum. 
The interpretation of the above expression is straightforward. 
In the absence of NP, the modified isotope shifts for $m$ transitions lie on a $(m-1)$-dimensional subspace. 
Consequently, $n=m+1$ independent isotope pairs (points on the generalized $m$-dimensional King plot), thus $m+2$ isotopes, are needed in order to experimentally test this prediction.
In cases where the system of equations is overconstrained, that is when there are more measurements than parameters to be determined, then one may consider fitting methods like those presented in Refs.~\cite{Frugiuele:2016rii,Solaro:2020dxz}. 

A given data set is planar if the volume of the $m$-dimensional parallelotope formed by the $m$ measured $\mnvec$ vectors (each of them being of $n$ dimensions, the number different isotope pairs $a_1,\ \dots,\ a_n$) is consistent with zero within experimental errors. 
Such a volume is given by the fully antisymmetric product of the measured $\mnvec$'s and the unit vector in isotope space $\Uhat=(1,\dots, 1)$,
\begin{align}
	\label{eq:volume}
	V
= 	\det(\mnvec_1,\dots,\mnvec_m,\Uhat)\,.
\end{align}
The volume induced by the theory ansatz in Eq.~\eqref{eq:linearrelation} is found to be 
\begin{align}
	\label{eq:volumetheory}
	V_{\rm th} 
= 	\frac{\alpNP}{(n\!-\!2)!}  
	\epsilon_{i_2\dots i_{n}}\epsilon_{a_1\dots a_n}1_{a_1}X_{i_2}
	h_{a_2}m\nu_{i_3}^{a_3}\dots \,m\nu_{i_{n}}^{a_n}\,,
\end{align}
with the transition indices $i_{1}\dots i_{n-1}$ and the isotope pairs indices $a_{1}\dots a_{n}$ are taking values in $1\dots n-1$ and $1\dots n$, respectively. 
Therefore, the NP coupling is given by
\begin{align}
	\label{eq:alphaNP}
	\alpNP
= 	\frac{V}{V_{\rm th}(\alpha_{\rm NP}=1)}\,.
\end{align}
We learn that even if the original notion of King linearity breaks down, NP can be probed. 
In the following we refer to this method as \textit{generalized King~(GK)} analysis.  

In order to understand the limitations of the GK method, we write $V_{\rm th}$ in terms of the different nuclear and electronic parameters
\begin{align}
	V_{\rm th}
=	\det\left( F_{il} , X_i\right) \det\left( h_a, m\mu_a, m\lambda_{l,a} \right) \, .
\end{align}
Therefore, we can see that if any of the two parameters are aligned, the NP sensitivity is lost.

Let us consider the case of 4 IS pairs and 3 transitions, \textit{i.e.} $n=4$ and $m=3$. 
The uncertainty in $\alpNP$ can be derived by usual first-order error propagation from Eq.~\eqref{eq:alphaNP}, $\sigma[\alpNP] = \sqrt{ \sum_k (\partial\alpNP/\partial \CO_k )^2\sigma[k]^2}\,$ where $k$ runs over all experimental observables $\CO_k$.
Under the assumption that only $\sigma[{\nu^a_i}]\ne0$, it can be written in terms of the 2D King plot NLs
\begin{align}
	\sigma[\alpNP]
=& 	\sqrt{ \sum_{a,l}\left( \frac{\sigma[\nu^a_i]}{\mu_a}  \sigma^{al}_{\alpNP} \right)^2 } \, ,  
\end{align}
with 
\begin{align}
	\sigma^{al}_{\alpNP}
=&	12 \frac{ \epsilon_{ijk} X_i \NL^{jk}_{\bar{a}} \, \epsilon_{a_1 a_2 a_3 a_4} h_{a_1} \NL^{\bar{l}}_{a_2 a_3 a_4} }
	{\left( \epsilon_{ijk} \epsilon_{a_1 a_2 a_3 a_4} X_i h_{a1} \NL^{jk}_{a_2 a_3 a_4} \right)^2} \, ,
\end{align}
where the $\bar{a}$ is a 3D vector in the isotope space excluding the isotope pair $a$, $\bar{l}$ is a 2D vector in the transition space excluding the transition $l$ and
\begin{align}
	\label{eq:2DNL}
	\NL^{ij}_{a_1 a_2 a_3} =(\mnvec_i \times \mnvec_j) \cdot \Uhat \, ,
\end{align}
with the vectors $\mnvec_i$ in the isotope space built from the isotope pairs $a_1$, $a_2$ and $a_3$.

Finally, we emphasize that the GK analysis above can be straightforwardly repeated in cases having a different combination of known and unknown nuclear spurions. Also, while here we assumed that the isotope mass uncertainties were irrelevant, that is nuclear masses were known with sufficient precision, the latter can always be replaced by adding one more measured transition as in the NMK analysis in Section~\ref{sec:masslessking}.

\vspace{0.25cm}

%%%%%%%%%%%%%%%%%%%%%%%%%%%%%%%
\section{Sensitivity to new physics: the Ytterbium case} 
\label{sec:Yb}
%%%%%%%%%%%%%%%%%%%%%%%%%%%%%%%

In this section, we demonstrate the potential of the NMK and GK analyses presented in Sections~\ref{sec:masslessking} and~\ref{sec:generalkingNP} in probing NP forces. 
We focus on the case of a new spin-independent interaction mediated by a force carrier $\phi$ of spin $s=0,1,2$. 
The effective Yukawa potential associated with the new intra-atomic force is given by $V_\phi(r) = -\alpNP (A-Z) e^{-m_\phi r}/r\,$, where $m_\phi$ is the force carrier mass and  $\alpNP = (-1)^s y_e y_n/4\pi\,$, with $y_e\,(y_n)$ is the $\phi$-electron\,($\phi$-neutron) coupling and $\gamma_a=A_0-A$.
Note that we gross over a potential $\phi$-proton coupling here, since its effect would  cancel in the IS. 

Systems with several (measurable) optical clock transitions are quite rare.
Ytterbium is a rather unique element where three such transitions have been observed: 
two in the Yb$^+$ ion, $4f^{14}6s\ (^2S_{1/2})\to 4f^{14}5d\ (^2D_{3/2})$ and $4f^{14}6s\ (^2S_{1/2})\to 4f^{13}6s^2\ (^2F_{7/2})$, and one in neutral Yb, $4f^{14}6s^2\ (^1S_0)\to 4f^{14}6s6p\ (^3P_0)$, at a wavelength of 436\,nm, 467\,nm and 578\,nm, respectively. The parenthesis indicates the term symbol of the angular momentum quantum number associated with the multi-electron state. For simplicity, in the following, we refer to the above transitions as $S-D_{3/2}$, $S-F$ and $S-P$, respectively.
The three of them have been measured with sub-Hz precision for the $A=171$ isotope~\cite{PTB,huntemann2014improved,Brown:2017lvs,Yb0Meas}. 
Another candidate clock transition in neutral Yb, yet to be measured with high precision, is $4f^{14}6s6p\ (^3P_0)\to 4f^{13}6s^25d\ J=2$  at 1696\,nm~\cite{Safronova:2018quw}. Moreover, ytterbium has five stable isotopes with zero nuclear spin, $A=168,170,172,174,176$ and one radioactive isotope ($A=166$) with a half-life of $\approx 2.4\,$days. 
Hence, in principle, generalized King linearity can be tested with very high precision in $3-$dimensional, or even 4-dimensional, King plots combining clock transitions in Yb$^0$ and in Yb$^+$.

The SM leading correction to Eq.~\eqref{eq:LOshift} in the Yb system is the second-order FS~\cite{Flambaum:2017onb}. 
By matching this correction to Eq.~\eqref{eq:ISgen}, it can be written as a single term, $F_{i2}\lambda_{2,a}$, with $\lambda_{2,a}\approx (\delr{a})^2$.
This correction contributes to the SM NLs of the usual King plot. 
We quantify the NL induced on the King plot as
\begin{align}
	\label{eq:RNL}
	\CR_\NL \equiv \frac{\NL^{ij}_{a_1a_2a_3}}{ \sigma[\NL^{ij}_{a_1a_2a_3}] }\,, 
\end{align}
where $\CR_\NL\gtrsim \CO(1)$ is a clear signal of King linearity breaking. 
The definition of Eq.~\eqref{eq:RNL} can be generalized to any type of King plot, see also Ref.~\cite{Solaro:2020dxz}.  

In our numerical estimations below, we evaluate the electronic parameters, $K_i$, $F_i$, $F_{i2}$~(the SM parameters) and $X_i(m_\phi)$ (the NP parameters as functions of $m_\phi$) by using configuration interaction~(CI) implemented in \ambit~\cite{KAHL2019232}, see Appendix~\ref{sec:nums} for details.
The resulting estimates of the SM electronic parameters are reported in Table~\ref{tab:elecparam}. 
The nuclear parameters, such as the nuclear masses~$(m_A)$ and the charge radius differences~$(\delr{a})$ are taken from~\cite{Wang_2017,NISTASD} and~\cite{ANGELI201369}, respectively, and reported in Table~\ref{tab:nuclearparam} for completeness. 

Then, we evaluate  $\CR^{\rm SM}_\NL$ for the three ytterbium transitions and a given choice of three isotope pairs formed by either of $A=168,\,170,\,172,\,176$ and the reference $A_0=174$.
In addition, we also estimate the NL between the $S\to D_{3/2}$ and $S\to D_{5/2}$ Yb$^+$ transitions that were measured in Ref.~\cite{Counts:2020aws}.
For simplicity, we assume a common error bar, $\sigma_\nu$, for all the transitions and neglect other sources of uncertainty, like nuclear mass uncertainty. 
The results are summarized in Table~\ref{tab:RSM}. In particular, we observe that $\CR^{\rm SM}_\NL\sim\CO(0.2-5)\,\kHz/\sigma_\nu$ from which  we conclude  that  NLs from subleading SM effects become important once the accuracy of the IS measurements crosses the kHz level.
%%%%%%%%%%%%%%%%%%%%%%%%%%%%%%%
\begin{table}[t]
\begin{center}
\begin{tabular}{ccccc}
\hline\hline
Transitions 
&  \multicolumn{4}{c}{$|\CR^{\rm SM}_\NL|(\kHz/\sigma_\nu)$} \\
\hline
Isotopes
& \tiny{170,172,176} 
& \tiny{168,172,176} 
& \tiny{168,170,176} 
& \tiny{168,170,172} \\
\hline
$S-F$, $S-D_{3/2}$
& 1.8 & 1.8 & 0.52 & 1.7\\
\hline
$S-P$, $S-D_{3/2}$
& 2.9 & 2.8 & 0.8  & 2.7 \\
\hline
$S-P$, $S-F$
& 4.6 & 4.4 & 1.3 & 4.2\\
\hline\hline
$S-D_{3/2}$, $S-D_{5/2}$ 
& 0.75 & 0.72 & 0.21 & 0.70\\
\hline\hline
\end{tabular}
\end{center}
\caption{The estimated SM 2D NL relative to the uncertainty, $\CR^{\rm SM}_\NL$ (see Eq.~\eqref{eq:RNL}), for different combination of transitions and isotopes with respect to $A_0=174$.}
\label{tab:RSM}
\end{table}%
%%%%%%%%%%%%%%%%%%%%%%%%%%%%%%%

Finally, we estimate the reach of the NMK and GK analyses to probe a new electron-neutron Yukawa interaction, and compare those to the original 2D King~(2DK) analysis with zero and finite isotope mass uncertainties. 
While bounds on NP resulting from an experimental IS analysis depend only on a minimal set of theory inputs, \textit{viz.} the NP parameters $X_i$ and $\gamma_a$, but not on the SM parameters, the estimated NP reach requires however knowledge of the electronic and nuclear parameters. In principle, the SM parameters can be extracted by fitting the data. Here, instead, we use the estimates in Table~\ref{tab:elecparam} for the electronic constants.
In constructing our estimates for the projected reach of the various King plot analyses, we consider the $S-F$ and $S-D_{3/2}$ narrow transitions in Yb$^+$, and we use the $S-P$ transition of Yb$^0$ whenever a third transition is needed. 
Concerning the isotopes, we choose $A_0=174$ and $A=170,\, 172, \, 176$ and  $A=168$ whenever the analysis requires a fourth isotope. 
Motivated by the experimental results of Ref.~\cite{Manovitz:2019czu}, we derive  projections under the assumption that the IS measurement accuracy for all transitions is $\sigma[{\nu^{a}_i}]= 0.01\,$Hz.

%%%%%%%%%%%%%%%%%%%%%%%%%%%%%%%
\begin{figure*}[t]
\begin{center}
\includegraphics[width=0.49\textwidth]{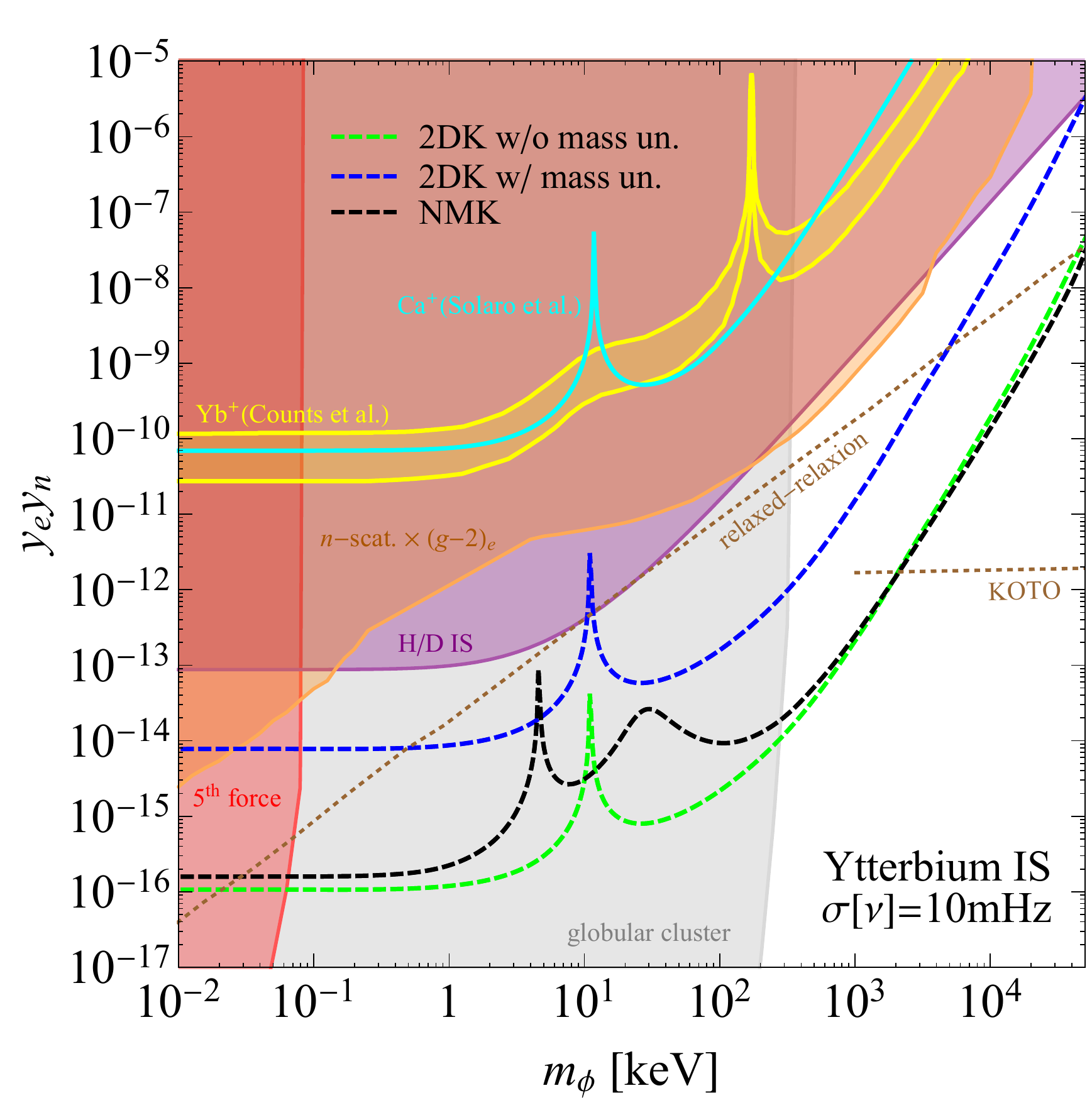}
\includegraphics[width=0.49\textwidth]{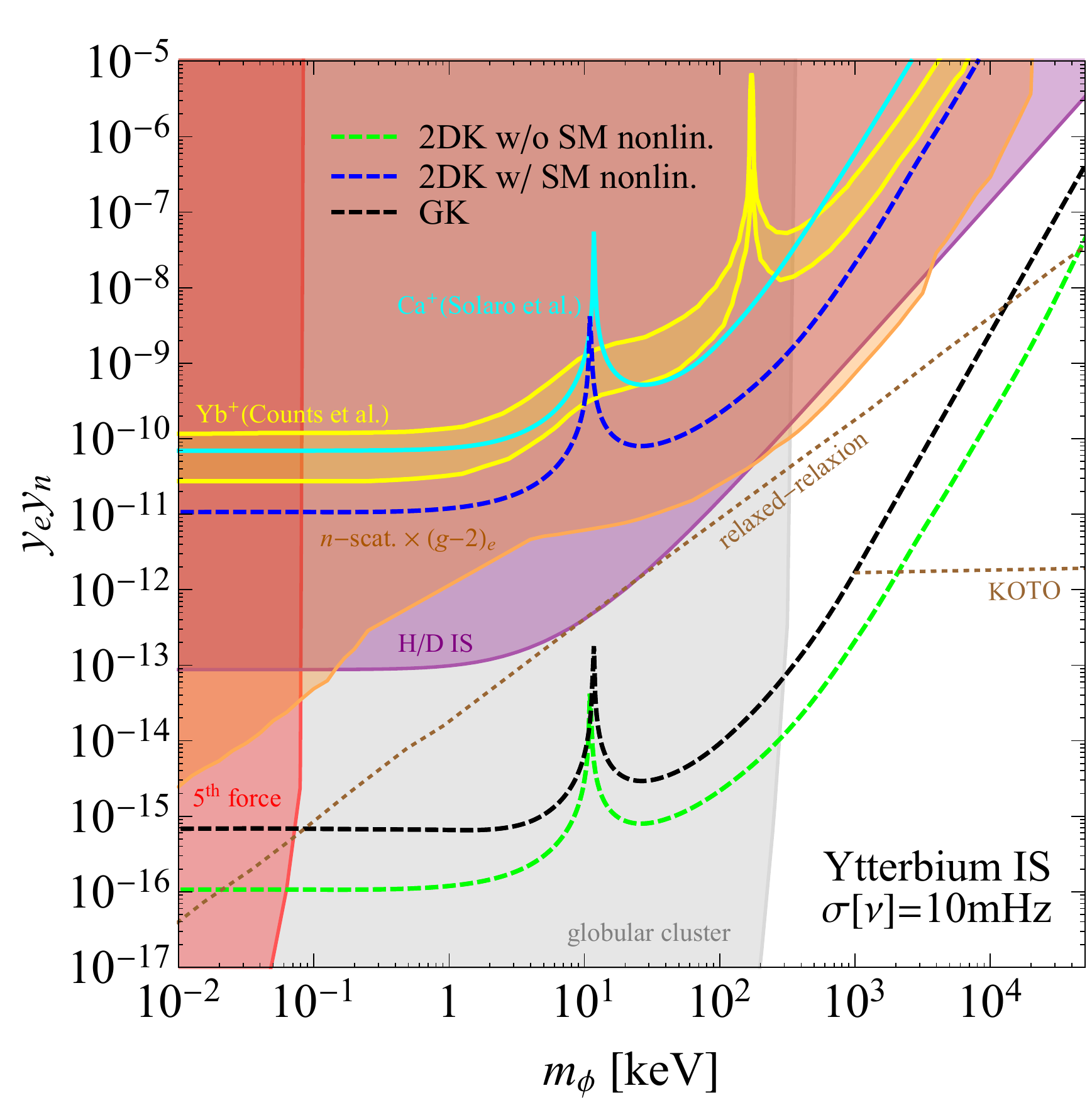}
\caption{
The projected IS bounds in the Yb system with a universal IS measurement uncertainty of $\sigma[\nu]=10\,$mHz (dashed lines). (left) Comparison of projections from 2D King~(2DK) analysis based on $S\to D_{3/2}$ and $S\to F$ transitions with zero non linearities~(NLs) from the SM and omitting (green) or including (blue) current nuclear mass uncertainties, and from the no-mass King analysis~(NMK) adding the $S\to P$ transition (black). 
(right) Comparison of projections from the 2DK analysis with zero nuclear mass uncertainty and including (blue) or not (green)  $\NL_{\rm SM}$ and from the generalized King analysis~(GK) adding the $S\to P$ transition (black).  
The existing NP bounds from IS spectroscopy in Yb$^+$ (the preferred 95\,\%~CL NP interval of Ref.~\cite{Counts:2020aws}) and Ca$^+$ (the upper bound from Ref.~\cite{Solaro:2020dxz}) are shown in yellow and cyan solid lines, respectively. 
Also shown are the current bounds from 5th force searches~\cite{Bordag:2001qi,Bordag:2009zzd}, electron-neutron scattering~\cite{Adler:1974ge}, neutron-nucleus scattering~\cite{Barbieri:1975xy,Leeb:1992qf,Nesvizhevsky:2007by,Pokotilovski:2006up} combined with the electron magnetic moment~\cite{Berengut:2017zuo}, $1S-2S$ hydrogen-deuterium (HD) IS~\cite{Delaunay:2017dku} and globular cluster~\cite{Hardy:2016kme}. 
The dotted lines indicate the relaxed-relaxion and a hypothetical scalar that may explain the KOTO results. 
}
\label{fig:Yb2D3DGen}
\end{center}
\end{figure*}
%%%%%%%%%%%%%%%%%%%%%%%%%%%%%%%

We perform two sets of comparisons.
The first set focuses on the effect of the nuclear mass uncertainties. Thus, in this case, we assume zero NLs from the SM, $\NL_{\rm SM}=0$.
We then compare the following three analyses:
\begin{enumerate}
\item the 2DK with $\sigma[m_A]=0$;
\item the 2DK with $\sigma[m_A]\ne0$;
\item and the NMK of Section~\ref{sec:masslessking}. 
\end{enumerate}

The second set focuses instead on dealing with large SM NLs, while neglecting nuclear mass uncertainties, that is we set $\sigma[{m_A}]=0$.  
We then compare the following three analyses:
\begin{enumerate}
\item the 2DK with $\NL_{\rm SM}=0$; 
\item the 2DK with $\NL_{\rm NP}\ll \NL_{\rm SM}\approx1.5\times 10^8\,[\MHz^2{\rm amu}^2]$; 
\item and the GK of Section~\ref{sec:generalkingNP}. 
\end{enumerate}
Figure~\ref{fig:Yb2D3DGen} shows the expected 95\,\% confidence level~(CL) bounds, \textit{i.e.} $\alpNP<2\sigma[\alpNP]$, of the above two sets of comparisons, along with the current IS bound from Ca$^+$~\cite{Solaro:2020dxz} and the 95\,\%~CL interval preferred by the NP interpretation of the Yb$^+$ results of Ref.~\cite{Counts:2020aws}. 
For sake of completeness, we also show existing bounds from 5th force searches~\cite{Bordag:2001qi,Bordag:2009zzd}, electron-neutron scattering~\cite{Adler:1974ge}, neutron-nucleus scattering~\cite{Barbieri:1975xy,Leeb:1992qf,Nesvizhevsky:2007by,Pokotilovski:2006up} combined with the electron magnetic moment~\cite{Berengut:2017zuo},  hydrogen-deuterium (HD) IS for the $1S-2S$ transition where the charge radius difference is determined by muonic spectroscopy (under the assumption of no NP in the muonic sector)~\cite{Delaunay:2017dku} and globular cluster~\cite{Hardy:2016kme}. 
As pointed out in~\cite{Bar:2019ifz} reasonable doubt persists regarding the SN\,1987A bound~\cite{Raffelt:2012sp}, therefore we chose to omit it. 
For reference, we also mark the theoretical lines predicted by the relaxed-relaxion model of Ref.~\cite{Banerjee:2020kww} and by an hypothetical scalar with a finite lifetime that can accommodate the recent KOTO results~\cite{Kitahara:2019lws,Banerjee:2020kww,KOTOslides}.

From the left panel of Fig.~\ref{fig:Yb2D3DGen} we learn that the nuclear mass uncertainties, if kept at the current level, will become an important limitation in searching for NP with the usual King plot analysis. 
However, this can be overcome by using instead the NMK analysis provided accurate IS measurements of an additional narrow transition are possible.
The NMK analysis shows  sensitivity to NP similar to the original 2DK with zero mass uncertainties. This motivates the use of a purely spectroscopy-based approach which does not rely on modified IS, but rather on IS frequencies directly.  

The right panel of Fig.~\ref{fig:Yb2D3DGen} demonstrates the importance of accounting for NLs emerging from the SM higher order contributions to the IS, on the one hand, and how it can potentially limit the sensitivity of the usual King plot analysis to probe NP, on the other hand. 
Nevertheless, as we argued in Section~\ref{sec:generalking}, the leading NL contributions of the SM can be fitted from the data and the GK analysis shows  sensitivity to NP that compares to the original 2DK analysis.

%%%%%%%%%%%%%%%%%%%%%%%%%%%%%%%
\section{Conclusions}
\label{sec:summary}
%%%%%%%%%%%%%%%%%%%%%%%%%%%%%%%

There have been some important developments in IS spectroscopy. In particular IS measurements recently reached an unprecedented level of accuracy of better than $10\,$mHz in strontium~\cite{Manovitz:2019czu}.  
This implies that the NP sensitivity with IS searches can potentially be improved by more than five orders of magnitude compared to the first NP-dedicated King analyses~\cite{Counts:2020aws,Solaro:2020dxz}.
However, once entering in such a new regime of precision, experiments will quickly face two limitations from 
(i)~the uncertainties in the nuclear mass determinations; and
(ii)~the SM contributions to the NL of the King plot. 

In this work, we present two data-driven methods to overcome them, namely the no-mass King analysis and the generalised King analysis, respectively. The proposed methods are based on the idea that more measurements of electronic transitions and IS allow us to determine additional nuclear and electronic parameters purely from the data without additional theory inputs. 
With enough transitions and isotope pairs, the above two methods can be combined into a single analysis which is insensitive to both the leading SM NL effect and the nuclear masses. 

Finally, while the constraining power of both methods were demonstrated with ytterbium transitions, both can be applied equally well to other systems where  the needed additional clock transitions may come from very different charged ion states~\cite{Micke2020}.

%%%%%%%%%%%%%%%%%%%%%%%%%%%%%%
\section*{Acknowledgment} 
%%%%%%%%%%%%%%%%%%%%%%%%%%%%%%

We thank V.V.~Flambaum, E.~Fuchs, R.~Ozeri, G.~Perez, S.~Schiller, Y.~Shpilman and C.~G.~Solaro for fruitful discussions.
JCB is supported in this work by the Australian Research Council (DP190100974).
CD is supported by the program Initiative d'Excellence of Grenoble-Alpes University under grant Contract Number ANR-15-IDEX-02.
YS is supported by the United States-Israel Binational Science Foundation~(BSF) (NSF-BSF program Grant No. 2018683) and by the Azrieli Foundation.
YS is Taub fellow (supported by the Taub Family Foundation). 
This work was performed in part at Aspen Center for Physics, which is supported by National Science Foundation grant PHY-1607611.

%%%%%%%%%%%%%%%%%%%%%%%%%%%%%%%
\appendix
%%%%%%%%%%%%%%%%%%%%%%%%%%%%%%%

%%%%%%%%%%%%%%%%%%%%%%%%%%%%%%%
\section{Details of the numerical calculations}
\label{sec:nums}
%%%%%%%%%%%%%%%%%%%%%%%%%%%%%%%
	
The numerical atomic calculations in this work were performed using \ambit~\cite{KAHL2019232}. 
For both neutral Yb and Yb$^+$, the calculation begins with Dirac-Fock (relativistic Hartee-Fock) to obtain the self-consistent potential and core orbitals up to and including $5s^2\ 5p^6\ 4f^{14}$. 
We then create the $6s$, $6p$, and $5d$ valence orbitals as eigenstates of the core Dirac-Fock potential. 
Higher energy orbitals are generated by successively multiplying the orbitals by the radial function $r$ and orthogonalizing to the lower core and valence orbitals~\cite{bogdanovich1983approximate}.
The $5f$ orbital is generated by multiplying the $5d$ orbital by $r$ and orthogonalizing to the $4f$ core orbital. 
This procedure saturates the configuration interaction~(CI) space more quickly than $B$ splines. 
We also found that the resulting values of $X_i$ are more stable in the high $m_\phi$ limit than when using $B$ splines, which we hypothesise may be due to small irregularities in the $B$ spline orbitals near the origin (see, \textit{e.g.},~\cite{BELOY2008310} for a discussion of this topic). 
For this reason we have not included MBPT in the current work (except to calculate the mass shift operators $K_i$, which are more sensitive to correlations and less sensitive to the wavefunction at the nucleus).
	
For the neutral Yb case we generate excited states up to $12spdf$. 
The configuration interaction includes all many-body basis states generated by single and double excitations from $6s^2$, $6s\,6p$, $6s\,5d$, and $6p^2$. 
We also allow single excitations from the core $5s$, $5p$, and $4f$ shells. 
The electronic parameters $X_i$ and $F_i$ are generated by adding the corresponding potential to the Hamiltonian and taking the difference.
	
The treatment of Yb$^+$ is similar, however since we are targeting the $4f^{13}\,6s^2$ configuration the CI matrix grows much more quickly than in the neutral Yb case. 
We include excited states up to $8spdf$ and take single and double excitations from the configurations $4f^{14}\,6s$, $4f^{14}\,6p$, $4f^{14}\,5d$, $4f^{13}\,6s^2$, $4f^{13}\,6s\,5d$, $4f^{13}\,6p^2$, and $4f^{13}\,5d^2$. 
We also limit the number of additional excitations from the $4f$ shell to one.

Our numerical estimation for the electronic parameters are given in Table~\ref{tab:elecparam}. 
In addition Table~\ref{tab:nuclearparam} contains the nuclear parameters relevant to this work. 

%%%%%%%%%%%%%%%%%%%%%%%%%%%%%%%
\begin{table}[htb]
\begin{center}
\begin{ruledtabular}
\begin{tabular}{cccc}
Transition & $K_i\,$[GHz.amu] & $F_i\,$[MHz/fm$^2$] & $F_{i2}\,$[MHz/fm$^4$] \\
\hline
$S-F$ & 5111 & 36218& -201.12\\
$S-D_{3/2}$ & -1040 &-14968 & 83.247\\
$S-D_{5/2}$ & -1247 &-14715& 81.908\\
$S-P$ &-655 & -9719.2 & 54.277\\
\end{tabular}
\end{ruledtabular}
\end{center}
\caption{The electronic parameters for the Yb and Yb$^+$ transitions relevant to this work, as estimated using the \ambit\ code, see text for details.}
\label{tab:elecparam}
\end{table}%
%%%%%%%%%%%%%%%%%%%%%%%%%%%%%%%

%%%%%%%%%%%%%%%%%%%%%%%%%%%%%%%
\begin{table}[htb]
\begin{center}
\begin{ruledtabular}
\begin{tabular}{cccc}
Isotope & $m_A$\,[amu] & $\sigma[m_A]$\,[$10^{-9}$amu] & $\delr{a}$[fm$^2$]  \\
\hline
$168$ & $167.8958990$ & $1300$ & -0.5406 \\
$170$ & $169.896777237$ & $11$ & -0.3845\\
$172$ & $171.898396650$ & $15$ & -0.2366\\
$174$ & $173.900877539$ & $12$ & -0.1159\\
$176$ & $175.904584700$ & $16$ & --\\
\end{tabular}
\end{ruledtabular}
\end{center}
\caption{The nuclear parameters of Yb relevant to this work. 
The nuclear masses $m_A=m_{A,{\rm Yb}^0}-70m_e+E_b$, with $E_b$ the binding energy, are taken from Ref.~\cite{Wang_2017,NISTASD}. Note that the mass uncertainty is only due to the uncertainty in the atomic mass $m_{A,{\rm Yb}^0}$ as the correlated errors on $m_e$ and $E_b$ are mostly canceled in $\mu_a$.
The difference in the rms charge radii are taken from Ref.~\cite{ANGELI201369} and are relative to $A=176$, we neglect the associated errors.}
\label{tab:nuclearparam}
\end{table}%
%%%%%%%%%%%%%%%%%%%%%%%%%%%%%%%

%%%%%%%%%%%%%%%%%%%%%%%%%%%%%%%%%%%%%%%%%%%%%%%%%%%%
\bibliographystyle{utphys}
\bibliography{King_bib}
%%%%%%%%%%%%%%%%%%%%%%%%%%%%%%%%%%%%%%%%%%%%%%%%%%%%
	
\end{document}